\documentclass[12pt]{article}
\usepackage{amsmath, amssymb, amscd, amsthm, bm, amsfonts,threeparttable,float}
\usepackage{multicol}
\usepackage{graphicx}
\usepackage{hyperref}
\usepackage{natbib}
\usepackage{subfigure}
\def\argmin{\mbox{argmin}}

\title{C-learning in estimation of optimal individualized treatment regimes for recurrent disease}
\author{Zi-Shu Zhan$^1$, Jin-Lun Zhang$^{1}$, Chen Shi$^1$,  Xiao-Han Xu$^1$, Chun-Quan Ou$^{1,*}$}
\date{}

\begin{document}

\maketitle

\begin{abstract}
Recurrent events, characterized by the repeated occurrence of the same event in an individual, are a common type of data in medical research. Motivated by cancer recurrences, we aim to estimate the optimal individualized treatment regime (ITR) that effectively mitigates such recurrent events. An ITR is a decision rule that assigns the optimal treatment to each patient, based on personalized information, with the aim of maximizing the overall therapeutic benefits. However, existing studies of estimating ITR mainly focus on first-time events rather than recurrent events. To address the issue of determining the optimal ITR for recurrent events, we propose the Recurrent C-learning (ReCL) method to identify the optimal ITR from two or multiple treatment options. The proposed method reformulates the optimization problem into a weighted classification problem. We introduce three estimators for the misclassification cost: the outcome regression estimator, the inverse probability weighting estimator, and the augmented inverse probability weighting estimator. The ReCL method leverages classification techniques to generate an interpretable optimal ITR tailored for recurrent event data. The advantages of the ReCL method are demonstrated through simulations under various scenarios. Furthermore, based on real data on colorectal cancer treatments, we employ this novel method to derive interpretable tree treatment regimes for colorectal cancer, thus providing a practical framework for enhancing treatment strategies.
\end{abstract}

\section{Introduction}\label{section-introduction}

\label{intro}
In recent years, the analysis of recurrent event data has increasingly gained the attention of researchers. Recurrent events, such as cancer recurrences, refer to multiple episodes of the same adverse event in an individual, which are commonly encountered in clinical trials and epidemiological studies. Traditional survival analyses, which focus solely on the time-to-first event and fail to fully utilize the recurrent data, do not adequately capture the complete progression of a disease in individuals. Many statistical methods have been developed to analyze recurrent event data, as comprehensively reviewed by \cite{cook2007} and \cite{amorim2015}. These models are mainly categorized into conditional models and marginal models. Conditional models mainly focus on modeling the intensity function \citep{ag1982,pwp1981,frailty2020}. In contrast, marginal models concentrate on the overall distribution of event times and the cumulative number of events \citep{lawless1995,lin1998,lin2001,sun2020}. Recently, there has been a growing focus on causal inference with recurrent event outcomes \citep{schaubel2010,su2022causal,furberg2023,janvin2024causal}. However, these methods are aimed at estimating the average treatment effect rather than the individual-level effect. 

In clinical settings, assigning a uniform treatment that is optimal on average to all patients may be suboptimal for some individuals as different subgroups may exhibit varied treatment responses. Comparatively, precision medicine seeks to identify an individualized treatment regime (ITR) that tailors treatments based on specific patient characteristics, thereby optimizing therapeutic outcomes or so-called value functions for all individuals \citep{kosorok2019,zhao2024}. Regression-based methods, such as Q-learning \citep{qian2011} and A-learning \citep{shi2018highdimensional}, are frequently used to estimate the optimal ITR but may yield suboptimal estimation due to model misspecifications. An alternative approach is directly optimizing the value function estimator. For example, \cite{zhang2012robust} proposed both the inverse probability weighting (IPW) estimator and the augmented inverse probability weighting (AIPW) estimator of the value function, which seek the optimal ITR by directly optimizing an estimator for the overall population mean outcome. Additionally, \cite{zhang2012class} proposed a C-learning method and \cite{zhao2012estimating} proposed an outcome weighting learning (OWL) method to estimate ITR from a classification perspective. Both of them reformulated the optimization problem of the value function estimator into a weighted classification challenge, and employed different classifiers to approximate the true optimal ITR. More developments and studies along this research line have been conducted on continuous outcome \citep{laber2015tree,zhou2017,gao2024}, binary outcome \citep{guo2021}, time-to-first event data \citep{fang2021,cui2024,zhou2023}, but not yet recurrent event data. Therefore, new developments are necessary to estimate the optimal ITR for recurrent diseases.

Estimating optimal ITR within the recurrent event data framework presents significant challenges. Directly applying existing ITR approaches based on the time-to-first event ignores all subsequent recurrent events, potentially biasing the evaluation of treatment efficacy \citep{amorim2015}. Moreover, the definition of the optimal treatment for recurrent events differs from that for the first event. Existing methods mainly focus on binary treatments. However, estimating optimal ITRs can be complicated when there are three or more treatment options \citep{Tao2017,zhou2018outcome,Xue:2021}. Additionally, the European Medicines Agency's qualification opinion on recurrent event analyses highlights the importance of interpretability of statistical results \citep{akacha2018}. An optimal ITR that considers recurrent events and offers clear interpretability is essential for facilitating decision-making in disease treatment and prognosis. 

In this study, we propose a Recurrent C-learning (ReCL) method, an extension of the C-learning framework \citep{zhang2012class} to recurrent event data, aiming at estimating the optimal ITR that minimizes mean number of recurrent events. The proposed method reformulates the optimization problem into a weighted classification problem. We introduce three estimators for the misclassification cost: the outcome regression (OR) estimator, the IPW estimator, and the AIPW estimator. Furthermore, we extend the proposed method to the multi-arm treatment issue. Specifically, we transform the optimization problem into an example-dependent, cost-sensitive classification problem \citep{Elkan:2001}. Powerful classification techniques, coupled with the data space expansion technique \citep{Abe:2004}, can be used to solve this problem and estimate the optimal ITR. The ReCL method retains the flexibility of C-learning and integrates both IPW and AIPW estimators based on pseudo-observations (PO) \citep{andersen2003}, offering robustness against model misspecification. In general, the ReCL method is expected to tackle challenges related to recurrent event data. It enhances estimation accuracy by integrating machine learning methods and also prioritizes the interpretability of treatment regimes using tree-based classifiers.

The remainder of this paper is organized as follows: Section \ref{method} introduces the mathematical framework and the estimation algorithm. Section \ref{simu} demonstrates the finite sample performance through numerical studies. We apply the novel method to the colorectal cancer readmission data in Section \ref{real}. A discussion is provided in Section \ref{conclu}. Theoretical results and additional numerical results are provided in the supplementary materials.

\section{Methods}
\label{method}
\subsection{Notations}
Consider a clinical trial or an observational study of a total maximum follow-up period $\tau$ with $n$ subjects. Let ${\bm X}\in\mathcal{X}$ denote $p$-dimensional vector of covariates and $A\in\mathcal{A}$ denote the treatment assigned. Under the treatment assignment $A=a$, we define $\tilde{N}^a(t)$ as the potential outcomes for the number of events that occurred by time $t$. In addition, we define $C$ and ${Y}(t)=I(C\ge t)$ as the right censoring time and the at-risk function, respectively. Instead of observing the potential counting process $\tilde{N}^a(t)$, we only observe $N(t)$. Then, the observed sample is $\{{\bm X}_i,A_i,N_i(t),C_i,Y_i(t)\}$. Similar to \cite{su2022causal}, we make the following assumptions: (i) No Undermeasured Confounder Assumption (NUCA): $\tilde{N}^a(\cdot)\perp A|{\bm X}, a\in \mathcal{A}$; (ii) the Stable Unit Treatment Value Assumption (SUTVA): each subject's potential outcomes are unaffected by others' treatments; there is no different form of each treatment level; (iii) censoring at random: $C\perp\tilde{N}^a(\cdot)|{\bm X}, a\in\mathcal{A}$; (iv) the random sample $\{{\bm X}_i,A_i,N_i(t),C_i,Y_i(t)\}$ is independent and identically distributed. Under the these assumptions, we have $N(t)=\sum_{a\in\mathcal{A}}I(A=a)\tilde{N}^a(t)$, 
where $a\in\mathcal{A}$. We are interested in the ITR $g\in\mathcal{G}$, a decision rule mapping $\mathcal{X}$ to $\mathcal{A}$. The potential counting process associated with regime $g$ is denoted as $\tilde{N}^g(t)$. Generally, the fewer the recurrence events, the better. Thus, for a given time $t$, the optimal ITR $g^{\rm opt}$ satisfies
\begin{align}\label{eq:obj1}
   g^{\rm opt}=\argmin_{g\in\mathcal{G}}E\{\tilde{N}^g(t)\}. 
\end{align}

\subsection{Recurrent C-learning for two treatments}
In this part, we adopt the C-learning proposed by \cite{zhang2012class} to accommodate recurrent events. We first consider the binary treatment setting with $\mathcal{A}=\{0,1\}$. Analogous to \cite{qian2011}, we define the $Q$-function at time $t$ as $Q^a(t,{\bm X})=E\{{N}(t)|{\bm X}, A = a\}$.
Based on the Q-function, define a contrast function $C(t, {\bm X})=Q^1(t, {\bm X})-Q^0(t,{\bm X})$ and $W=I[C(t,{\bm X})<0]$, we can show that $g^{\rm opt}$ minimizes an expected weighted misclassification error as follows:
\begin{align}\label{eq:class}
    g^{\rm opt}=\argmin_{g\in\mathcal{G}}E\{|C(t,{\bm X})|I(g\neq W)\}.
\end{align}
Specifically, we can refer to $|C(t,{\bm X})|$ as the misclassification cost and $W$ as the label. Details on the proof of the equivalence of equation (\ref{eq:obj1}) and (\ref{eq:class}) can be found in the supplementary materials. Thus, the optimal treatment regime at time $t$ can estimated by 
\begin{align}\label{eq:est1}
    \hat{g}^{\rm opt}=\argmin_{g\in\mathcal{G}}\frac{1}{n}\sum_{i=1}^n\left\{|\hat{C}(t,{\bm X}_i)|I(g\neq \hat{W}_i)\right\},
\end{align}
where $\hat{C}(t,{\bm X}_i)$ is the estimator of $C(t,{\bm X})$ for subject $i$ and $\hat{W}_i=I[\hat{C}(t, {\bm X}_i)<0]$. Here, we propose three different estimators for $C(t, {\bm X})$ and ${W}$: the OR estimator, the IPW estimator, and the AIPW estimator based on pseudo-observations.

\subsubsection{OR estimator}
Define $r^*(t,{\bm X}_i,A_i)$ as the posited model for $E[dN_i(t)|{\bm X}_i, A_i]$ and $\mu^*(t,{\bm X}_i,A_i)$ as the corresponding model for $E[N_i(t)|{\bm X}_i, A_i]$ based on $r^*(t,{\bm X}_i,A_i)$. For a given $t$, consider the following semiparametric multiplicative rate (SMR) model
\begin{align}\label{eq:smr}
    r^*(t,{\bm X}_i,A_i)=\exp\{{\bm X}_i^{\top}{\bm\alpha}+A_i{\bm X}_i^{\top}{\bm\beta}\}d\mu(t),
\end{align}
where ${\bm\alpha}$ and ${\bm\beta}$ are corresponding parameters, and $\mu(t)$ is the unspecified baseline mean function. More details about the above model can be found in \cite{lin2000}. 
Thus, under model (\ref{eq:smr}) one can estimate $C(t,{\bm X}_i)$ as follows:
\begin{align*}
\hat{C}^{\rm SMR}(t,{\bm X}_i) &=\hat{\mu}^*(t,{\bm X}_i,1)-\hat{\mu}^*(t,{\bm X}_i,0)\\
    &=\exp\{{\bm X}_i^{\top}\hat{\bm\alpha}+{\bm X}_i^{\top}\hat{\bm\beta}\}\hat{\mu}(t)-\exp\{{\bm X}_i^{\top}\hat{\bm\alpha}\}\hat{\mu}(t),
\end{align*}
where $\hat{\mu}^*(t,{\bm X}_i,a)$ is the estimator for ${\mu}^*(t,{\bm X}_i,a)$, $a\in\mathcal{A}$, and $\hat{\bm \alpha}$, $\hat{\bm \beta}$ and $\hat{\mu}(t)$ are estimators of ${\bm\alpha}$, ${\bm\beta}$ and ${\mu}(t)$, respectively. We can obtain $\hat{\bm \alpha}$, $\hat{\bm \beta}$ and $\hat{\mu}(t)$ by using the estimation equations proposed by \cite{cai2004}. We then obtain the corresponding $\hat{W}_{i}^{\rm SMR}=I[\hat{C}^{\rm SMR}(t,{\bm X}_i)<0]$. Note that the validity of the OR estimator is contingent upon the correct specification of the model (\ref{eq:smr}).

\subsubsection{IPW estimator}

Denote $\Lambda^a(t)=E[\tilde{N}^a(t)]$. When the event rate is unconditional on the event history, we refer to $\Lambda^a(t)$ as the cumulative rate function (CRF). Here, we construct the IPW estimator by employing the pseudo-observations method based on the CRF. Specifically, for a given $t$ and subject $i$, the pseudo-observation of CRF can be obtained by $\hat{\Lambda}_i(t)=n\hat{\Lambda}_{\rm NA}(t)-(n-1)\hat{\Lambda}_{\rm NA}^{-i}(t)$, where $\hat{\Lambda}_{\rm NA}=\sum_{i=1}^n\int_0^t1/\sum_{j=1}^nY_j(s)dN_i(s)$ is a Nelson-Aalen type estimator and $\hat{\Lambda}_{\rm NA}^{-i}(t)$ is the same
estimator without subject $i$. Thus, we have that 
\begin{align*}
E[\hat{\Lambda}_i(t)|{\bm X}_i,A_i]\approx E[N_i(t)|{\bm X}_i,A_i].
\end{align*}
Further information about $\hat{\Lambda}_i(t)$ can be found in \cite{su2022causal}. Define $\pi^*(a,{\bm X}_i)$ as the assumed model for the propensity score (PS) function $P(A_i=a|{\bm X}_i)$, $a\in\mathcal{A}$. Then, if $\pi^*(a,{\bm X}_i)$ is correctly specified, we can prove that 
\begin{align}\label{IPWprov}
    E\left[\frac{I(A_i=a)\hat{\Lambda}_i(t)}{\pi^*(a,{\bm X}_i)}\bigg|{\bm X}_i\right]\approx E\left[{N_i(t)}|{\bm X}_i,A_i=a\right],
\end{align}
The reason for (\ref{IPWprov}) is displayed in the supplementary materials.
And we can estimate $C(t,{\bm X}_i)$ as follows:
\begin{align*}
    \hat{C}^{\rm IPW}(t,{\bm X}_i)=\frac{I(A_i=1)\hat{\Lambda}_i(t)}{\hat{\pi}^*(1,{\bm X}_i)}-\frac{I(A_i=0)\hat{\Lambda}_i(t)}{\hat{\pi}^*(0,{\bm X}_i)},
\end{align*}
where $\hat{\pi}^*(a,{\bm X}_i), a\in\mathcal{A}$ is the estimator of the PS function based on the assumed model $\pi^*(a,{\bm X}_i)$, and we can obtain the corresponding $\hat{W}^{\rm IPW}_i=I[\hat{C}^{\rm IPW}(t,{\bm X}_i)<0]$.

\subsubsection{AIPW estimator}
As widely discussed in the literature \citep{rosenbaum1983,zhang2012robust}, incorrect preliminary models for $E[N_i(t)|{\bm X}_i, A_i]$ and ${P}(A_i=a|{\bm X}_i)$ can lead to suboptimal results. Fortunately, when either $\mu^*(t,{\bm X}_i,A_i)$ or $\pi^*(a,{\bm X}_i)$ is correctly specified, we can prove that
\begin{align}\label{AIPWprov}
    E\left\{\frac{I(A_i=a)\hat{\Lambda}_i(t)}{\pi^*(a,{\bm X}_i)}+\left[1-\frac{I(A_i=a)}{\pi^*(a,{\bm X}_i)}\right]\mu^*(t, {\bm X}_i, a)\bigg |{\bm X}_i\right\}=E[N_i(t)|{\bm X}_i,A_i=a].
\end{align}
Therefore, we consider the doubly robust AIPW estimators by combining the OR estimator and the IPW estimator as follows:
\begin{align*}
\hat{C}^{\rm AIPW}(t,{\bm X}_i)&=\left\{\frac{I(A_i=1) \hat{\Lambda}_i(t)}{\hat{\pi}^*(1,{\bm X}_i)}+\left[1-\frac{I(A_i=1)}{\hat{\pi}^*(1,{\bm X}_i)}\right]\hat{\mu}^*(t,{\bm X}_i,1)\right\} \\
  &~~~~ - \left\{\frac{I(A_i=0) \hat{\Lambda}_i(t)}{\hat{\pi}^*(0,{\bm X}_i)}+\left[1-\frac{I(A_i=0)}{\hat{\pi}^*(0,{\bm X}_i)}\right]\hat{\mu}^*(t,{\bm X}_i,0)\right\},
\end{align*}
and $\hat{W}^{\rm AIPW}_i=I[\hat{C}^{\rm AIPW}(t,{\bm X}_i)<0]$. This AIPW estimator is robust to model misspecification in the sense that estimates remain consistent if either the OR model or the PS model, but not both, is misspecified.

\subsection{Extension of recurrent C-learning to
multiple treatments}
In this part, we expand the binary treatment space to include $K$ treatments, that is,  $\mathcal{A}=\{1,\cdots,K\}$. Recall that the Q-function corresponding to $A=1,\cdots,K$ treatments are $Q^1(t,{\bm X})$,\\$\cdots,Q^K(t,{\bm X})$, respectively. We denote the $k$-th order statistics of the Q-function as $Q^{(k)}(t,{\bm X})$; then, we have $Q^{(1)}(t,{\bm X})\le\cdots\le Q^{(K)}(t,{\bm X})$. Let $l^{(k)}(t,{\bm X})$, $1\le k\le K$ be the corresponding treatment of $Q^{(k)}(t,{\bm X})$. Therefore, the contrast function can be defined as $C^k(t,{\bm X})=Q^k(t,{\bm X})-Q^{(1)}(t,{\bm X})$. Analogous to the binary case, the optimal treatment regime at time $t$ can be expressed by 
\begin{align}\label{eq:obj2}
  {g}^{\rm opt}=\argmin_{g\in\mathcal{G}}E\left\{{C}^g(t,{\bm X})I[g\neq l^{(1)}(t,{\bm X})]\right\}.
\end{align}
More details of the derivation can be found in the supplementary materials. Accordingly, this leads to an estimation problem for $g^{\rm opt}$ as follows:
\begin{align}\label{eq:est2}
    \hat{g}^{\rm opt}=\argmin_{g\in\mathcal{G}}\frac{1}{n}\sum_{i=1}^n\left\{\hat{C}^g(t,{\bm X}_i)I[g\neq \hat{l}^{(1)}(t,{\bm X}_i)]\right\},
\end{align}
where $\hat{C}^g(t,{\bm X}_i)$ is the estimator of $C^g(t,{\bm X})$ and $\hat{l}^{(1)}(t,{\bm X}_i)$ is the estimator of  $l^{(1)}(t,{\bm X}_i)$ for subject $i$, respectively. 
Correspondingly, we can obtain the OR estimator, IPW estimator and AIPW estimator for the case with multiple treatments as follows:
\begin{align*}
    \hat{C}^{g,\rm SMR}(t,{\bm X}_i)&=\hat{\mu}^{*}(t,{\bm X}_i,g)-\min_{a\in\mathcal{A}}\hat{\mu}^{*}(t,{\bm X}_i,a),\\
    \hat{C}^{g,\rm IPW}(t,{\bm X}_i)&=\frac{I(A_i=g)\hat{\Lambda}_i(t)}{\hat{\pi}^*(g,{\bm X}_i)}-\min_{a\in\mathcal{A}}\frac{I(A_i=a)\hat{\Lambda}_i(t)}{\hat{\pi}^*(a,{\bm X}_i)},\\
    \hat{C}^{g,\rm AIPW}(t,{\bm X}_i)&=\left\{\frac{I(A_i=g) \hat{\Lambda}_i(t)}{\hat{\pi}^*(g,{\bm X}_i)}+\left[1-\frac{I(A_i=g)}{\hat{\pi}^*(g,{\bm X}_i)}\right]\hat{\mu}^{*}(t,{\bm X}_i,g)\right\} \\
  &~~~~ - \min_{a\in\mathcal{A}}\left\{\frac{I(A_i=a) \hat{\Lambda}_i(t)}{\hat{\pi}^*(a,{\bm X}_i)}+\left[1-\frac{I(A_i=a)}{\hat{\pi}^*(a,{\bm X}_i)}\right]\hat{\mu}^*(t,{\bm X}_i,a)\right\},\\
  {\text{and}}~~~~~~~~~~~~~~~~~~~~~~~~&\\
  \hat{l}^{(1),\rm SMR}(t,{\bm X}_i)&=\argmin_{a\in\mathcal{A}}\hat{\mu}^{*}(t,{\bm X}_i,a),\\
    \hat{l}^{(1),\rm IPW}(t,{\bm X}_i)&=\argmin_{a\in\mathcal{A}}\frac{I(A_i=a)\hat{\Lambda}_i(t)}{\hat{\pi}^*(a,{\bm X}_i)},\\
    \hat{l}^{(1),\rm AIPW}(t,{\bm X}_i)&=\argmin_{a\in\mathcal{A}}\left\{\frac{I(A_i=a) \hat{\Lambda}_i(t)}{\hat{\pi}^*(a,{\bm X}_i)}+\left[1-\frac{I(A_i=a)}{\hat{\pi}^*(a,{\bm X}_i)}\right]\hat{\mu}^{*}(t,{\bm X}_i,a)\right\}.
\end{align*}

\subsection{Computation}
In practice, for binary treatments and multiple treatments, the PS function can be estimated using binary and multinomial logistic models. We also recommend using machine learning methods such as random forest for estimating the PS function. An alternative approach is SuperLearner algorithm \citep{van2007super}, which combines candidate parametric and nonparametric methods to identify the optimal combination that minimizes cross-validated risk. The SuperLearner has been shown to perform asymptotically as well as, or better than,  any individual candidate learner included in its ensemble. 
For the recurrent outcome model, we opt for the SMR model in the following numerical study for its simplicity. While other complicated or flexible methods, such as recurrent random forest \citep{loe2023random,murris2024}, could also be adopted and analyzed in a similar way; however, this is not the focus of our work. 
We adopt the classification and regression tree (CART) as the classifier. Hence, the estimated individualized treatment regime is a tree-type regime, which possesses considerable advantages in interpretability.

Note that the classification problem (\ref{eq:est2}) is an example-dependent, cost-sensitive, multi-class weighted classification issue. For a given $t$, let $\{{\bm X}_i, \hat{\bm C}_i(t)\}_{i=1}^n$ denote the classification problem, where $\hat{\bm C}_i(t)=(\hat{C}^1(t,{\bm X}_i),\cdots,\hat{C}^K(t,{\bm X}_i))$
is the cost vector at time $t$ for subject $i$. According to the definition of $\hat{C}^k(t,{\bm X}_i), k=1,\cdots,K, i = 1,\cdots, n$, the cost corresponding to $\hat{l}^{(1)}(t,{\bm X}_i)$ is zero. Thus we can derive the label of each patient from the cost vector. 
We use the data space expansion method to solve the classification problem $\{{\bm X}_i, \hat{\bm C}_i(t)\}_{i=1}^n$.
We first define 
\begin{eqnarray*}
\hat{U}_{i}^k=\hat{C}^k(t,{\bm X}_i)-\min_{1\leq s \leq K}\hat{C}^s(t,{\bm X}_i),~ k=1,\cdots, K.
\end{eqnarray*} 
Then, we modify the cost-sensitive weighted classification problem $\{{\bm X}_i, \hat{\bm C}_i(t)\}_{i=1}^n$ into a regular weighted classification problem $\mathop{\cup}\limits_{k=1}^K\{{\bm X}_i, k, \hat{U}_{i}^k\}_{i=1}^{n}$, where $k$ is the label and $\hat{U}_{i}^k$ is the corresponding misclassification error. The sample size of the data set after the expansion is $K$ times the original one.
As proven by \cite{Abe:2004}, solving the classification problem after data space expansion is equivalent to solving the original unmodified problem. Any existing classifier such as the above-mentioned CART, can be used to solve the modified classification problem.

\section{Simulation Designs}
\label{simu}
In this section, we conduct numerical studies to assess the performance of the proposed ReCL algorithm. 

\noindent{\bf Scenario 1. (Binary treatments)}\\
We first generate three covariates $(X_{1},X_{2},X_3)$ from the normal distribution $N(0,1)$ independently. We considered observational studies with the treatment assignment depending on covariates. Thus, treatment $A$ assigned to each patient is randomly generated from Bernoulli distribution with PS function $P(A=1|{\bm X})=\exp(0.3X_1-0.5X_2)/[1+\exp(0.3X_1-0.5X_2)]$. This yields approximately 50\% of subjects receiving $A=1$ treatment. The optimal regime is defined as 
\begin{align*}
g^{\rm opt}=I(X_1>-1)I(X_2>-0.5).
\end{align*}
The recurrent event times are generated from the homogeneous Poisson process with rate function 
$$
E[dN(t)|{\bm X}, A]=\exp\{X_2+|1.5X_1-0.5|(A-g^{\rm opt})^2-0.8\}d\mu(t),
$$
where $\mu(t)=0.5t$. A study duration of $\tau = 4$ is employed, and the censoring time for each subject is independently generated
from a Uniform($\tau-1$, $\tau$), which is also independent of the event processes. In Scenario 1, we experimentally investigate the performance of the following methods at the pre-specified time point $t=2,3$:

\begin{itemize}
    \item {ReCL-AIPW-T}: the proposed AIPW-based ReCL method with the correct specified PS model;
    \item {ReCL-AIPW-F}: the proposed AIPW-based ReCL method with the misspecified PS model as ``$A\sim X_1+Z$, where $Z=\exp(X_3)$";
    \item {ReCL-IPW}: the proposed IPW-based ReCL method with the correct specified PS model;
    \item {ReCL-SMR}: the proposed SMR-based ReCL method;
    \item {First}: the C-learning method for only considering the first-time event \citep{Zhan2024};
    \item {Random}: the method with random allocation of treatments; 
    \item {Optimal}: the method of treatment allocation according to $g^{\rm opt}$.
\end{itemize}
 \noindent{We simulate 100 replicates with sample size $n = 400, 600$.}\\

\noindent{\bf Scenario 2. (Multi-arm treatment)} \\
In the second scenario, we generalize the binary treatment case to the multi-arm case. Based on covariates $(X_1, X_2, X_3)$ generated independently from $N(0,1)$, treatment $A$ is randomly generated according to the distribution $P(A=1)=\pi_1/\pi_{\rm sum}$, $P(A=2)=\pi_2/\pi_{\rm sum}$, $P(A=3)=\pi_{3}/\pi_{\rm sum}$, where $\pi_1=1, \pi_2=\exp(X_1-X_2), \pi_3=\exp(0.5X_1-X_2)$ and $\pi_{\rm sum}=\sum_{k=1}^3\pi_k$. This yields approximately 40\% of subjects receiving $A=1$ treatment, 30\% of subjects receiving $A=2$ and the remaining 30\% of subjects receiving $A=3$. In this scenario, the optimal regime is defined as 
\begin{align*}
g^{\rm opt}=I(X_1>-0.5)[I(X_2>-0.5)+I(X_2>0.5)].
\end{align*}
The recurrent event times are generated from the homogeneous Poisson process with rate function 
$$
E[dN(t)|{\bm X}, A]=\exp\{0.3|1.5X_1-0.5|(A-g^{\rm opt})^2-0.3\}d\mu(t).
$$
We consider sample size $n=600, 800$. All other data-generating settings are the same as in Scenario 1. In Scenario 2, we similarly assess the performance of the methods mentioned earlier in Scenario 1.


When evaluating the proposed methods, the estimated value, defined as the mean number of recurrent events following the estimated ITR, is of the most interest. If the estimated ITR is satisfactory, the estimated value should be close to the true optimal value. Moreover, we show the assignment accuracy rate, say, the rate of treatments assigned by the estimated optimal ITR consistent with the true optimal treatments. A higher accuracy rate means better estimation of optimal ITRs. For assessing the generalization ability of the estimated optimal ITR, the accuracy rate and estimated value are computed using a testing set with a sample size of 5000. 

\subsection{Simulation results}
The results in Scenario 1 are presented in Figures \ref{fig:binAC} and \ref{fig:binev}, and the results in Scenario 2 are shown in Figures S1 and S2 in the supplementary materials.

For Scenario 1, the accuracy rate generally increases and the estimated value (the mean number of recurrent events) decreases as the sample size grows. In particular, the ReCL-AIPW-T method outperforms the others, with achieving the smallest average number of events and an assignment accuracy being larger than 0.9, which is close to the optimal level.
We use the SMR model to estimate the Q-function, thus the outcome model is always misspecified. However, in all cases, the ReCL-AIPW-T estimator outperforms the ReCL-IPW estimator, reflecting the robustness of the AIPW method to outcome model misspecification. When the PS model is correctly specified, the AIPW method (ReCL-AIPW-T) yields superior results compared to using an incorrect propensity score (ReCL-AIPW-F). It indicated that the results may be less promising when both the PS and outcome models are incorrectly specified. To our surprise, the ReCL-AIPW-F method, although not excelling in accuracy, demonstrates better performance in reducing the average number of recurrence events compared to ReCL-IPW. This may be because the ReCL-AIPW-F method incorporates information from both the outcome model and the PS model, even though this information may not be entirely accurate.

Overall, with Optimal excluded, the proposed method ReCL-AIPW-T are observed to perform the best in improving accuracy rates and reducing the mean number of recurrent events. As anticipated, the SMR method proves to be less competitive due to the model misspecification. We also observe that the C-learning method focusing solely on the first event performed poorly, indicating that overlooking subsequent recurrences can introduce bias in the estimation when our goal is to minimize the average number of recurrence events.

In Scenario 2, the performance of the ReCL-AIPW-T method remains commendable, with the lowest average number of events and the highest assignment accuracy. 
It is worth noting that, though the accuracy rate of ReCL-AIPW-T in Scenario 2 is lower than that in Scenario 1, the estimated mean number of value events is still close to the optimal one. Given the constraints of sample size, ReCL-AIPW-T may not always assign the optimal treatment to every individual. However, the outcomes under the estimated treatment regime are still favorable. In this scenario, the primary goal of the optimal ITR is to minimize the mean number of recurrence events, not necessarily to maximize the assignment accuracy. The minimization of the outcome of interest on average can be achieved at the expense of the assignment accuracy for those patients with relatively minor treatment effect differences. This trade-off reflects the inherent challenges of cost-sensitive, example-dependent, multi-class classification problems. Specifically, in the context of the limited sample size, the classification accuracy for samples with lower misclassification loss may be compromised.


\begin{figure}[H]
    \centering
    \includegraphics[width=14cm,height=15cm]{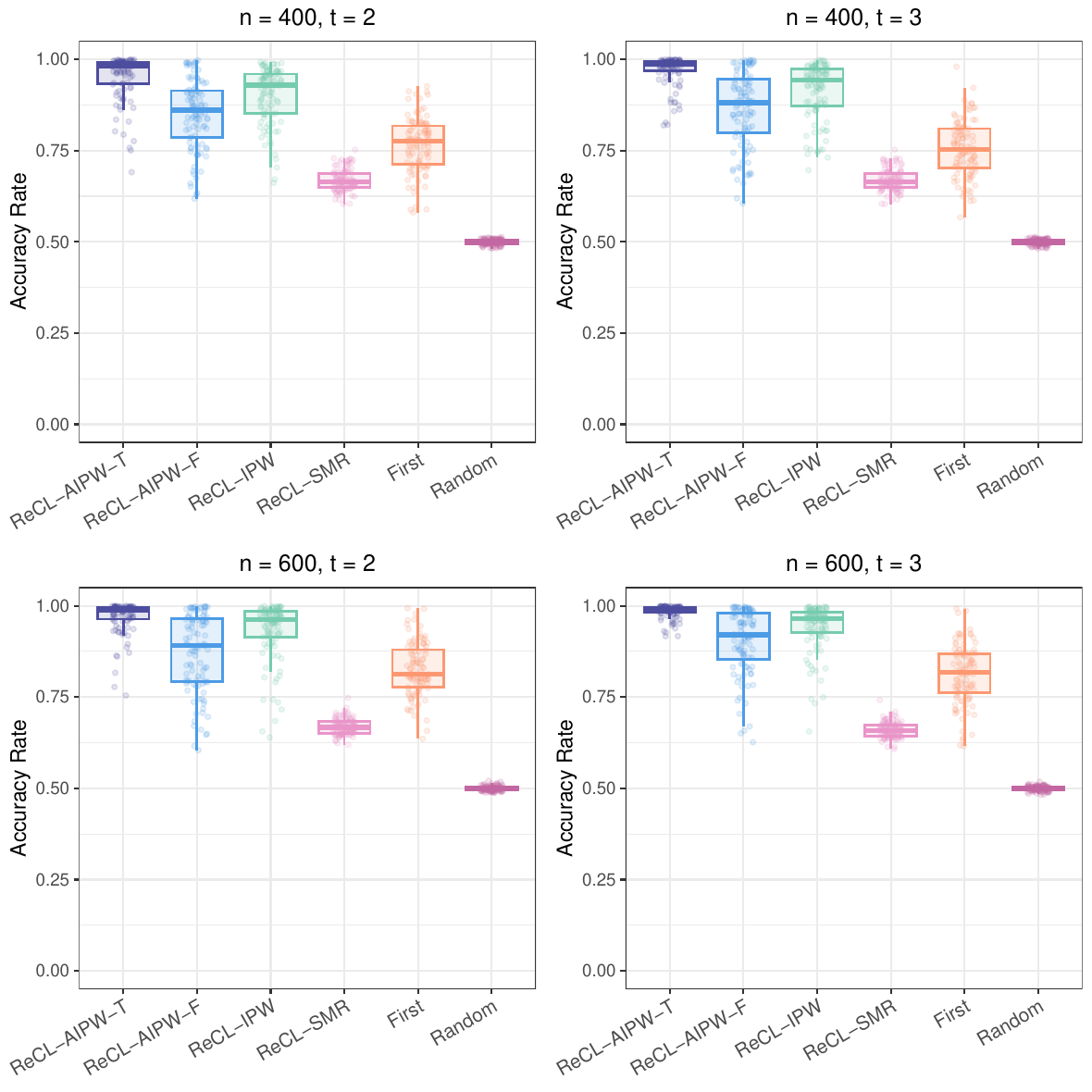}
    \caption{Simulation results of the accuracy rate in Scenario 1. Abbreviations: ReCL, Recurrent C-learning; AIPW, augmented inverse probability weighting; IPW, inverse probability weighting; T: correct specified propensity score model; F: misspecified propensity score model; SMR: semiparametric multiplicative model.}
    \label{fig:binAC}
\end{figure}

\begin{figure}[H]
    \centering
    \includegraphics[width=16cm]{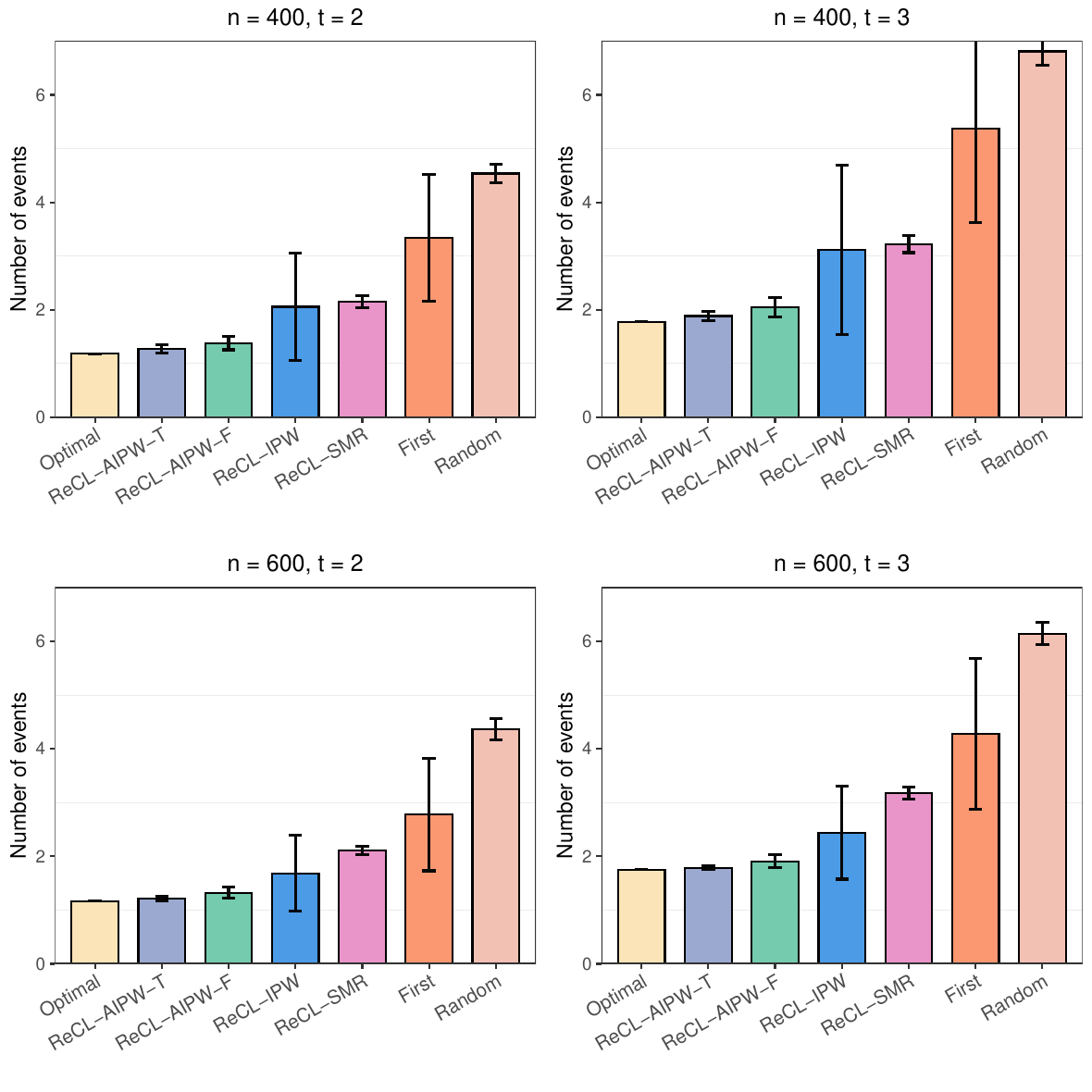}
    \caption{Simulation results of the estimated mean number of recurrent events with standard error in Scenario 1. Abbreviations: ReCL, Recurrent C-learning; AIPW, augmented inverse probability weighting; IPW, inverse probability weighting; T: correct specified propensity score model; F: misspecified propensity score model; SMR: semiparametric multiplicative model.}
    \label{fig:binev}
\end{figure}

\section{Real data analysis}
\label{real}
We apply the proposed methodology to a dataset concerning hospital readmissions of colorectal cancer patients from Hospital de Bellvitge in Barcelona, Spain.
This data set is accessible through the \texttt{frailtypack} package \citep{frailty} in R, which primarily serves to evaluate the effectiveness of chemotherapy, in the treatment of colorectal cancer. 
More details about the data can be found in \cite{gonzalez2005}. The primary outcome
of interest is the number of rehospitalizations following surgery in patients diagnosed with colorectal cancer during a pre-specified time point. Each of the 403 patients was followed up over time, with the maximum follow-up duration being 2176 days, and each readmission was documented. After the initial hospital admission due to colorectal cancer, 199 out of the 403 individuals experienced no readmissions, 150 (37.22\%) had one or two readmissions, and the remaining 54 (13.40\%) had three to 22 readmissions. Some patients received chemotherapy, a decision that may have been influenced by factors such as sex (0: male, 1: female) and Dukes' tumor stage (categorized into three groups: 1: stages A–B, 2: stage C, and 3: stage D).

We apply the proposed methods (ReCL-AIPW, ReCL-IPW and ReCL-SMR) to estimate the ITR and determine the optimal treatment for each individual. For comparison, we also implement the First and Random methods discussed in the simulation study. Based on the alignment between actual treatments and estimated optimal treatments, all individuals can be divided into two groups: one consisting of individuals who received treatments concordant with the estimated optimal recommendations (Concordance), and another comprising the remaining individuals (Disconcordance). We leverage SMR to fit the Q-function and SuperLearner to estimate the propensity scores on sex and Dukes' tumor stage for comparative methods that require such estimations. The pre-specified time points $t$ are set at 316 and 2176, corresponding to the one-third quantile of observed time and the maximum observed time, respectively.
Figure S3 in the supplementary materials shows the concordance pattern of the observed treatments and the estimated optimal treatments. We observe that the proposed methods (ReCL-AIPW and ReCL-IPW) have the highest proportion of concordance with the observed treatments at various time points, and these two methods yield similar treatment selection results, particularly when the set time $t=2176$.  

In real data analysis, since the true model generating the data is unknown, we use empirical values, similar to the idea of \cite{ye2023}, to estimate the average number of recurrences corresponding to treatments estimated by different methods as follows:
\begin{align*}
    \frac{\sum_{i=1}^n\frac{\hat{\Lambda}_i(t)I(A_i=\hat{g})}{\hat{\pi}^*(A_i,{\bm X}_i)}}{\sum_{i=1}^n\frac{I(A_i=\hat{g})}{\hat{\pi}^*(A_i,{\bm X}_i)}},
\end{align*}
where $\hat{g}$ is the estimated treatment of different methods, and $\hat{\pi}^*(A_i,{\bm X}_i)$ is the estimated propensity score based on SuperLearner. 
We list the estimated outcomes with $t=316$ or $2176$ in Table 1, where a smaller value indicates better model performance. It can be seen that our ReCL-AIPW/ReCL-IPW methods achieve the overall best performance in terms of reducing the number of recurrence events following the
suggested ITRs. For instance, when $t=2176$, the average number of recurrences corresponding to ReCL-AIPW and ReCL-IPW are both 2.90, while the average number of recurrences in the observed actual data is 4.16. In Figure S3 in the supplementary materials, approximately 44\% of the individuals receive treatments that differ from ReCL-AIPW and ReCL-IPW's recommendations; If all individuals use the recommended treatments, it could be expected that the number of recurrence events would be reduced by 1.26 on average.

\begin{table}[H]
    \centering
    \begin{threeparttable}[b]
    \caption{Results for real data analysis: The empirical mean number of recurrence corresponding to treatment of the observed data and
the estimated ITRs of different methods.}
    \begin{tabular}{ccccccc}
    \hline
       $t$ (Days)  & Observed & ReCL-AIPW & ReCL-IPW & ReCL-SMR & First & Random \\
       \hline
       316  & 0.59 & 0.48 & 0.47 & 0.49 & 0.49 & 0.73\\
       2176 & 4.16 & 2.90 & 2.90 & 5.59 & 4.88 & 3.48\\
       \hline
    \end{tabular}
    \begin{tablenotes}
    \small
    \item Abbreviations: Observed, the observed treatment in the real data; ReCL, Recurrent C-learning; AIPW, augmented inverse probability weighting; IPW, inverse probability weighting; SMR: semiparametric multiplicative model.
    \end{tablenotes}
    \end{threeparttable}
    \label{real:table}
\end{table}

To further evaluate the advantages of the proposed methods, we plot the estimated CRFs for both concordant and disconcordant groups after adjusting for the confounders in Figure \ref{real:CRF}. More details on the estimation of CRFs can be found in the supplementary materials. Overall, the ReCL-IPW and ReCL-AIPW methods outperform other methods. When $t=316$, the results of ReCL-AIPW, ReCL-IPW, ReCL-SMR and First are not significantly different, but all indicate that individuals whose actual treatments are concordant with the treatments recommended by these methods tend to experience fewer cumulative number of readmissions compared to their disconcordant counterparts. When the pre-specified time is set early, since the number of recurrences among patients is generally low, focusing only on the first recurrence and the average number of recurrences yields similar results. 
Besides, following the Random method’s recommendations increases the cumulative number of readmissions, indicating its inferiority to the current physician’s strategy. When the pre-specified time $t=2176$, adherence to the treatment recommendations from ReCL-AIPW and ReCL-IPW can reduce the number of recurrences. However, the advantages of treatments recommended by the methods of ReCL-SMR and First no longer exist when the timeframe exceeds 2000 days.  

\begin{figure}[H]
    \centering
    \begin{minipage}{0.49\linewidth}
    \centerline{\includegraphics[width=3.5in]{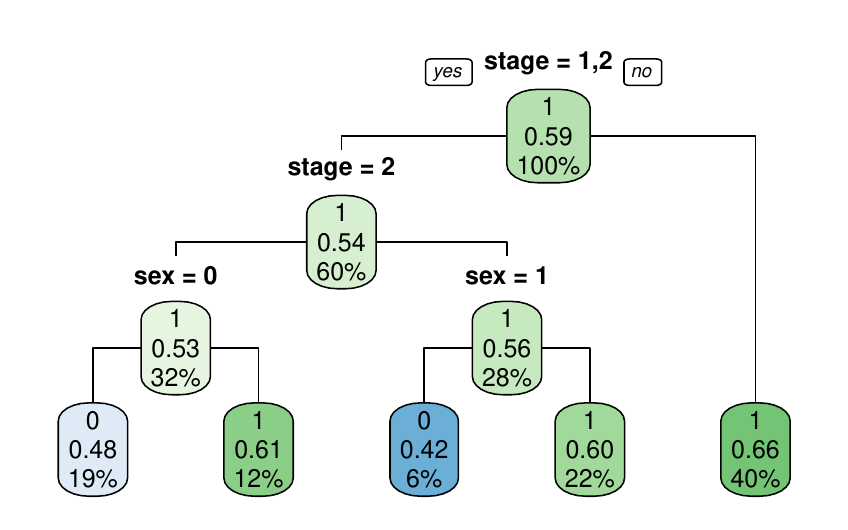}}
    \centerline{(a) $t=316$ (Days)}
    \end{minipage}
    \begin{minipage}{0.5\linewidth}
    \centerline{\includegraphics[width=3.45in]{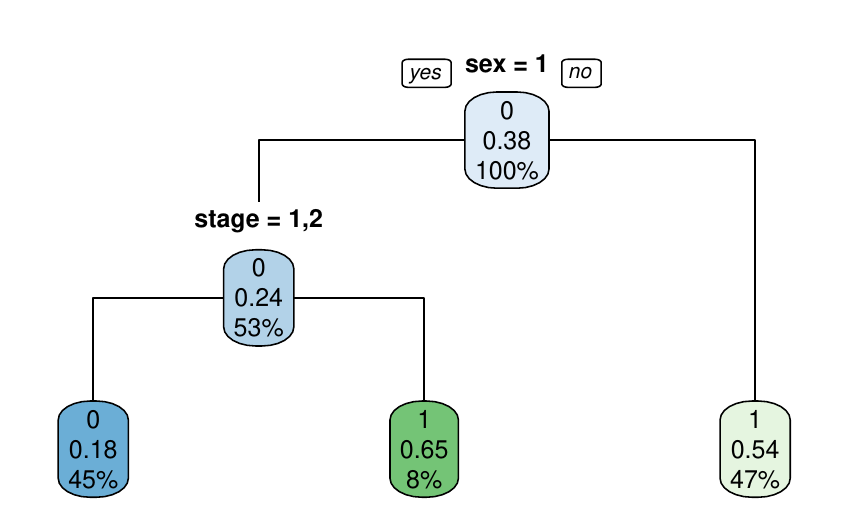}}
    \centerline{(b) $t=2176$ (Days)}
    \end{minipage}
    
    \caption{ Results for real data analysis: The estimated optimal ITR of ReCL-IPW at each stage. Treatment: 1, chemotherapy; 0, non-chemotherapy; sex: 0, male; 1, female; stage (Dukes' tumoral stage): 1, stages A–B; 2, stage C; 3, stage D. Abbreviations: ReCL, Reccurent C-learning; IPW, inverse probability weighting.}
    \label{real:tree}
\end{figure}

Given the superiority of our proposed methods over the others, we utilized these methods to draw decision tree diagrams. Since ReCL-AIPW and ReCL-IPW perform similarly in the real data analysis, we exemplify with ReCL-IPW to develop the decision tree for the optimal ITR of colorectal cancer. The direction of each tree branch is dictated by decision points, collectively shaping the strategy's final action. The estimated ITRs are depicted in Figure \ref{real:tree}. This tree-based decision-making provides intuitive interpretability. In each box in Figure \ref{real:tree}, the first row describes the estimated optimal treatment (1: chemotherapy; 0: non-chemotherapy), the second row displays the estimated probability of treatment 1, and the third row indicates the proportion of individuals in the node.

In Figure \ref{real:tree} (a), starting from the root, conditions like stage=1, 2 guide the path through the left (satisfied) or the right (unsatisfied) nodes, assigning treatment choices. This structured approach mirrors clinical decision-making, facilitating comprehension. For instance, within $t=316$, for patients with metastatic colorectal cancer (stage=3), chemotherapy treatment is recommended. As reported by \cite{cervantes2023}, chemotherapy after surgery (i.e. adjuvant chemotherapy) can be used to kill any remaining cancer cells and help reduce the risk of recurrence. Clinicians often present adjuvant chemotherapy to patients as ‘taking out an insurance policy’. \cite{hernandez2023} emphasize that the standard care for metastatic colorectal cancer primarily involves surgical resection and systemic therapy including chemotherapy. However, the costs are substantial and the benefits of adjuvant chemotherapy are not guaranteed if colorectal cancer is in its early stages. Therefore, it is necessary to decide on adjuvant therapy based on the patient's individual circumstances \citep{yang2024}. These treatment recommendations, as shown in our estimated tree diagram, demonstrate the clinical relevance and potential value of our methods to some extent.

\begin{figure}[H]
    \centering
    \subfigure[$t = 316$ (Days)]{
        \includegraphics[width=6in]{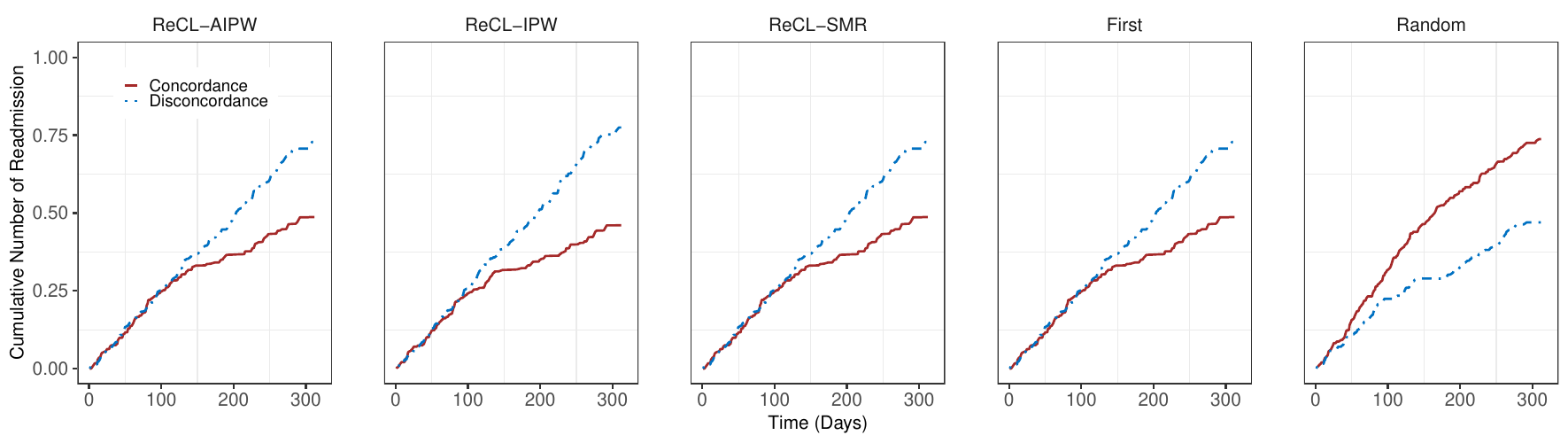}
    }
    \subfigure[$t = 2176$ (Days)]{
	\includegraphics[width=6in]{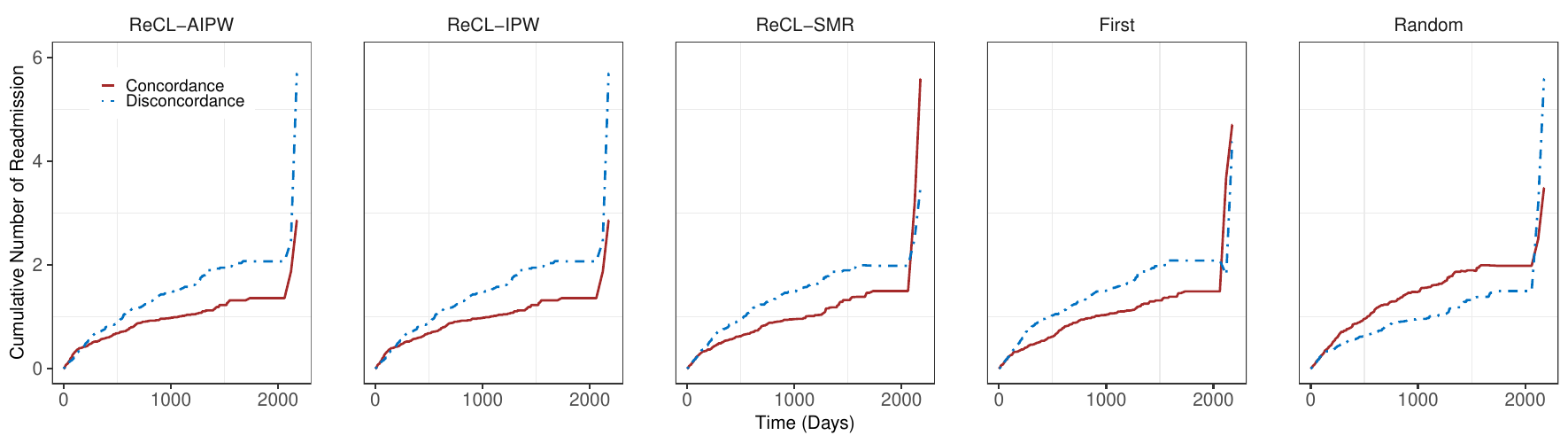}
    }
    \caption{Results for real data analysis: The estimated cumulative rate functions for individuals
whose truly received treatments are concordant/disconcordant with recommendations.  Abbreviations: ReCL, Recurrent C-learning; AIPW, augmented inverse probability weighting; IPW, inverse probability weighting; SMR: semiparametric multiplicative model.}
    \label{real:CRF}
\end{figure}

\section{Discussion}
\label{conclu}
The ReCL method we proposed extends the C-learning framework to accommodate recurrent event data. This method estimates the optimal ITR by minimizing recurrent event counts, demonstrating power and flexibility. We recast the minimization of average recurrences into managing an expected weighted misclassification error, introducing three estimators in the process: the OR estimator and two robust estimators (i.e. IPW and AIPW) that utilize pseudo-observations. The ITR derived from ReCL is both concise and highly interpretable, facilitating innovative diagnostics and therapeutic guidance. We adopt a tree-based strategy to mirror the clinical decision-making process and enhance physician understanding. 
Simulation results indicate that ReCL-AIPW, with the correct PS model, excels across diverse settings for both binary and multiple treatment scenarios. Overall, the two-treatment setup performs better. Given the cost-sensitive and sample-dependent properties of the multi-treatment problem, it is sometimes necessary to compromise by assigning suboptimal treatments to some individuals who exhibit lower misclassification rates.

Several improvements and extensions are worth future investigation. In long-term follow-up studies of recurrent events, non-ignorable terminal events such as death can complicate the analysis of personalized treatment strategies for recurrent events, as they terminate the recurrence process of diseases like colorectal cancer. Therefore, correcting the bias introduced by these terminal events in estimating optimal ITR is a crucial area for future research. Our current assumptions include that the censoring times are unaffected by treatments and are conditionally independent of the potential counting processes given covariates. However, these assumptions might not hold in practice, and the issue of informative censoring should be considered, as described in \cite{wang2001}. This paper focuses on a single-stage decision problem, but patients with cancer or recurrent diseases often undergo a prolonged sequence of initial treatments, recurrences, and salvage treatments \citep{huang2014}. Thus, it is important to construct a dynamic treatment regime (DTR) that utilizes all longitudinal data collected during the multi-stage process of disease recurrences and treatments, to minimize patients’ overall recurrent events. The development of optimal DTR within the survival data framework has been studied \citep{Goldberg:2012,Cho:2023}. Additionally, \cite{huang2014} provides a methodology for optimizing DTRs for recurrent diseases, aiming to maximize survival time potentially affected by post-recurrence salvage treatments. Hence, the study of a DTR estimator by minimizing the recurrent events as a direct criterion warrants further exploration.
\bibliographystyle{apalike}
\bibliography{references} 

\end{document}